\documentclass[prb,aps,showpacs,floatfix,floats,nobalancelastpage,twocolumn]{revtex4}
\usepackage{graphicx}
\usepackage{amsmath}
\usepackage{color}
\usepackage{dcolumn}
\usepackage{bm}
\usepackage{epstopdf}

\def\etal{~\textit{et~al.}}
\def\ra{\rangle}
\def\la{\langle}
\def\up{\uparrow}
\def\dn{\downarrow}

\def\ET{{$\kappa$-(ET)$_2$Cu$_2$(CN)$_3$}}

\begin{document}

\title{Effects of impurities in Spin Bose-Metal phase on a
two-leg triangular strip}
\author{Hsin-Hua Lai}
\affiliation{Department of Physics, California Institute of Technology, Pasadena, CA 91125}
\author{Olexei I. Motrunich}
\affiliation{Department of Physics, California Institute of Technology, Pasadena, CA 91125}
\date{\today}

\begin{abstract}
We study effects of nonmagnetic impurities in a Spin Bose-Metal (SBM)
phase discovered in a two-leg triangular strip spin-1/2 model with ring
exchanges (D.~N.~Sheng\etal, arXiv:0902.4210).
This phase is a quasi-1D descendant of a 2D spin liquid with spinon
Fermi sea, and the present study aims at interpolating between the
1D and 2D cases.
Different types of defects can be treated as local energy perturbations,
which we find are always relevant.  As a result, a nonmagnetic impurity
generically cuts the system into two decoupled parts.
We calculate bond energy and local spin susceptibility near the defect,
both of which can be measured in experiments.
The Spin Bose-Metal has dominant correlations at characteristic
incommensurate wavevectors that are revealed near the defect.
Thus, the bond energy shows a static texture oscillating as a function of
distance from the defect and decaying as a slow power law.
The local spin susceptibility also oscillates and actually {\it increases}
as a function of distance from the defect, similar to the effect
found in the 1D chain [S.~Eggert and I.~Affleck,
Phys.~Rev.~Lett. {\bf 75}, 934 (1995)].  We calculate the corresponding
power law exponents for the textures as a function of one Luttinger
parameter of the SBM theory.
\end{abstract}
\maketitle

\section{Introduction}

There has been much interest in spin liquid phases and much progress
has been made in our theoretical understanding of these
(see Ref.~\onlinecite{LeeNagaosaWen} for a review).
However, only recently several experimental candidates have emerged.
Among these, the triangular lattice based organic compound \ET\ shows
strong evidence of a gapless spin liquid.
\cite{Shimizu03, Kurosaki05, SYamashita08, MYamashita09}
One proposed theoretical state has a Fermi surface of fermionic spinons.
This appears as a good variational state\cite{ringxch} for an appropriate
spin model with ring exchanges and also as an appealing state in a
slave particle study\cite{SSLee} of the Hubbard model near the
Mott transition, leading to a U(1) gauge theory description.

The variational study is not sufficient to prove that a given state
is realized in the system, and the 2D gauge theory does not give
reliable information about the long-distance behavior.
Driven by the need for a controlled theoretical access to such phases,
Ref.~\onlinecite{Sheng09} considered the Heisenberg plus ring model on a
two-leg triangular strip and found a ladder descendant of the 2D
spin liquid in a wide regime of parameters; it also developed a
Bosonization description of the quasi-1D state.

The present work is motivated by $^{13}$C NMR experiments
\cite{Kawamoto04, Shimizu06, Kawamoto06, Itou08} in the organic
spin liquid material that observed strong inhomogeneous line broadening
at low temperatures.
Theoretical Ref.~\onlinecite{Gregor08} studied effects of nonmagnetic
impurities in the candidate spin liquid with spinon Fermi surface and
calculated the local spin susceptibility using mean field approach.
The susceptibility has an oscillating $2k_F$ component decaying
with a $1/x$ power law envelope.
A more complete gauge theory treatment is expected to modify this power
law,\cite{Kim03, Kolezhuk06, Gregor08} but one cannot calculate the
exponent quantitatively.

For comparison, the 1D Heisenberg chain can be loosely viewed as a 1D
version of the spinon Fermi sea state,\cite{KimLee}
and in this case the staggered
component of the local susceptibility {\it grows} away from an
impurity as $x^{1/2}$ in the limit of zero temperature and zero field.
This was discovered by Eggert and Affleck\cite{Eggert92, Eggert95} and is
responsible for strong inhomogeneous line broadening observed in
several 1D spin-1/2 chain materials.\cite{Takigawa, Fujiwara}

In this paper we calculate effects of nonmagnetic impurities in the
two-leg ladder descendant of the spin liquid using analytical approaches
developed in Ref.~\onlinecite{Sheng09}, in the hope of obtaining some
interpolation between the 1D chain and 2D spin liquid.
We find strong enhancement of the $2k_F$ components of the local
susceptibility compared with the mean field.  The susceptibility
increases away from an impurity as $\sim x^{1/2 - g/4} \geq x^{1/4}$,
where $g$ is one Luttinger parameter describing the phase\cite{Sheng09}
and can take values $g<1$.  This is a slower increase than in the
1D chain, but is still a dramatic effect. We also calculate bond textures
around the defect.

\section{Non-magnetic impurities in the Spin Bose-Metal on the ladder}

The spin system resides on the two-leg triangular ladder shown in
Fig.~\ref{fig:impurity}, which we can also view as a zigzag chain.
Throughout we assume that the model is in the described descendant
phase, which we will refer to as ``Spin Bose-Metal'' (SBM) following
Ref.~\onlinecite{Sheng09}.
Examples of non-magnetic defects are shown in Fig.~\ref{fig:impurity}
and are discussed in detail later.  Generally speaking, even though
there are different types of defects, we find that they eventually
(at low energies) cut the system into finite sections with essentially
open boundary conditions.\cite{Kane92, Eggert92, Eggert95}
We can then perform analytical calculations in a semi-infinite system
studying physical properties as a function of the distance from the
boundary.  In the following, we focus on induced textures in two
measurable quantities -- the bond energy and local spin susceptibility.
The physics is that an impurity perturbation has components on all
wavevectors and can directly ``nucleate'' the dominant bond energy
correlations.  The impurity also allows the uniform external magnetic
field to couple to the dominant spin correlations, producing textures
in the local susceptibility.

Following the description in Ref.~\onlinecite{Sheng09}, there are
three gapless modes with the fixed-point Lagrangian density
\begin{eqnarray}
\label{LSBM}
{\cal L}_{\rm SBM} &=& \frac{1}{2\pi g}
\left[\frac{1}{v} \left( \partial_\tau \theta_{\rho-} \right)^2
      + v \left( \partial_x \theta_{\rho-} \right)^2 \right] \\
\nonumber &+& \sum_{a=1,2} \frac{1}{2\pi}
\left[\frac{1}{v_a} \left( \partial_\tau \theta_{a\sigma} \right)^2
      + v_a \left( \partial_x \theta_{a\sigma} \right)^2 \right] ~.
\end{eqnarray}
Schematically, one route to this theory\cite{Sheng09} is via a
Bosonization treatment of electrons at half-filling on the ladder,
where we start with two bands, $a=1,2$, and assume that the
umklapp gaps out only the overall charge mode $\theta_{\rho+}$
while the other three modes $\theta_{\rho-}$, $\theta_{1\sigma}$,
and $\theta_{2\sigma}$ remain gapless. Note that in this paper, 
we simply postulate SBM phase and do not discuss how to stabilize it. 
However, our intuition is that with long-ranged repulsive interaction 
between electrons, we can make (relatively stable) C2S2 metallic phase 
(with four gapless modes $\rho_{+}, \rho_{-}, 1\sigma, 2\sigma$) go to 
the SBM phase, which is C1S2 Mott insulator with the overall charge mode $\rho+$ 
gapped by appropriate umklapp process. In addition, $g_{1 \sigma}$ and $g_{2 \sigma}$ 
are equal to 1 because of SU(2) spin invariance.

Ref.~\onlinecite{Sheng09} describes various observables in the SBM.
For the magnetic susceptibility calculations, we will need the
spin operator.  The $S^z$ component under Bosonization is
\begin{eqnarray}
S^z(x) \simeq
\frac{\partial_x (\theta_{1\sigma} + \theta_{2\sigma})}{\sqrt{2} \pi}
+ \sum_Q S^z_Q(x) e^{i Q x} ~.
\label{Sz}
\end{eqnarray}
The most important wavevectors are $Q = \pm 2k_{F1}$, $\pm 2k_{F2}$,
$\mp (k_{F1} + k_{F2}) = \pm \pi/2$, and $\pi$.
Each term can be expressed as in Ref.~\onlinecite{Sheng09}:
\begin{eqnarray}
S^z_{2k_{Fa}} &=& - e^{i\theta_{\rho+}} e^{\pm i \theta_{\rho-}}
\sin(\sqrt{2} \theta_{a\sigma}) ~, \\
S^z_{\pi/2} &=& e^{-i\theta_{\rho+}}
\Big[
- i \eta_{1\up} \eta_{2\up} e^{-i \theta_{\sigma+}}
\sin(\varphi_{\rho-} + \varphi_{\sigma-}) \\
\nonumber
&&~~~~~~~~~
+ i \eta_{1\dn} \eta_{2\dn} e^{i \theta_{\sigma+}}
\sin(\varphi_{\rho-} - \varphi_{\sigma-})
\Big] ~, \\
S^z_\pi &=&
[ \alpha \sin(2\theta_{\sigma+}) + \alpha' \sin(2\theta_{\sigma-}) ]
\sin(2\theta_{\rho+}) ~.
\end{eqnarray}
Throughout, we keep $\theta_{\rho+}$ general, but it is understood
to be pinned; details about the pinning value as well as the
Klein factors $\eta_{a\sigma}$ can be found in Ref.~\onlinecite{Sheng09}.
In the first line, the upper or lower sign corresponds to $a=1$ or $2$.
We also introduce combinations
$\theta_{\sigma\pm} = (\theta_{1\sigma} \pm \theta_{2\sigma})/\sqrt{2}$
and similarly for the conjugate fields $\varphi_{\sigma\pm}$.
In the last line, $\alpha$ and $\alpha'$ are independent numerical
constants.

When discussing non-magnetic defects and also in the bond energy texture
calculations, we need $n$-th neighbor bond energy operator like
\begin{equation}
B^{(n)}(x) \equiv \vec{S}_x \cdot \vec{S}_{x+n} ~.
\end{equation}
The bosonized form can be obtained from Ref.~\onlinecite{Sheng09}:
\begin{eqnarray}
B^{(n)} \simeq
\sum_{a=1,2} B^{(n)}_{2k_{Fa}} + B^{(n)}_{4k_{F1}} + B^{(n)}_{\pi/2} ~,
\label{Bn}
\end{eqnarray}
where we keep only the most important wavevectors and
\begin{eqnarray}
\label{bondenergy1}
&& B_{2k_{Fa}}^{(n)}(x) \sim \cos{(\sqrt{2} \theta_{a \sigma})}  \\
\nonumber && \hspace{1.5cm} \times
\cos{(2k_{Fa}x + \gamma^{(n)}_{2k_{Fa}} + \frac{\pi}{2} + \theta_{\rho+} \pm \theta_{\rho-})} ~, \\
\label{bondenergy2}
&& B_{4k_{F1}}^{(n)}(x) \sim
\cos{(4k_{F1}x + \gamma^{(n)}_{4k_{F1}} + 2\theta_{\rho+} + 2\theta_{\rho-})} ~, \\
\label{bondenergy3}
&& B_{\pi/2}^{(n)}(x) \sim \\
\nonumber &&
-i \eta_{1\up} \eta_{2\up} \cos{(\frac{\pi}{2}x + \gamma^{(n)}_{\frac{\pi}{2}} - \theta_{\rho+} - \theta_{\sigma+})} \sin{(\varphi_{\rho-} + \varphi_{\sigma-})} \\
\nonumber &&
-i \eta_{1\dn} \eta_{2\dn} \cos{(\frac{\pi}{2}x + \gamma^{(n)}_{\frac{\pi}{2}} - \theta_{\rho+} + \theta_{\sigma+})} \sin{(\varphi_{\rho-} - \varphi_{\sigma-})} .
\end{eqnarray}
We do not show real factors in front of all terms.
Here $\gamma^{(n)}_Q$ are phases that depend on $Q$ and the bond type:
\begin{eqnarray}
\gamma^{(n)}_Q = n Q/2 ~,
\end{eqnarray}
valid for $Q \neq \pi$.
Note also that since $4k_{F2} = -4k_{F1} \mod 2\pi$, there is only
one independent term $B_{4k_{F1}}$.

\subsection{Nonmagnetic defects treated as perturbations}

\begin{figure}[t]
   \centering
   \includegraphics[width=3.3in]{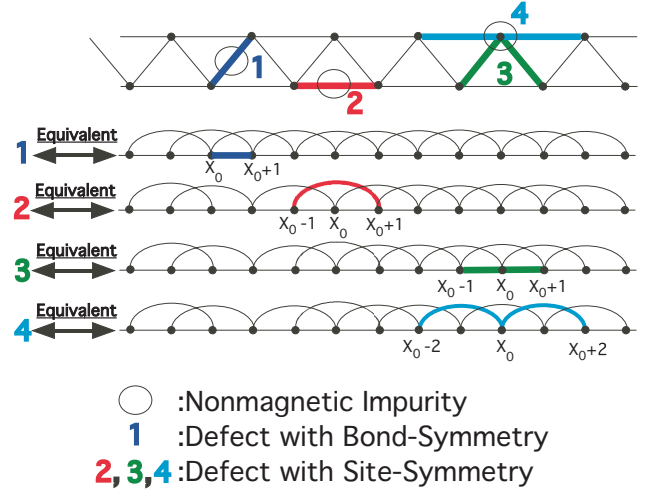}
   \caption{The top figure represents the original 2-leg triangular
ladder model with ring exchanges, and the thick lines represent the
defects due to the impurities.  The bottom figures represent the
corresponding defects in the equivalent 1D model.\cite{Sheng09}
1) represents the defect symmetric with respect to a bond center,
while 2), 3), and 4) represent defects symmetric about a site of the
1D chain. In general, different impurities will lead to different fixed points.
Impurity 1) will likely lead to a fixed point with decoupled
semi-infinite systems and a non-magnetic cluster containing an
even number of sites, while impurities 2), 3), and 4) will likely
lead to a fixed point with decoupled semi-infinite systems and an
effective spin formed by a cluster with an odd number of sites.}
   \label{fig:impurity}
\end{figure}

When a nonmagnetic defect is introduced at $x_0$, we can treat it as a
local perturbation in the Hamiltonian.\cite{Eggert92, Kane92}
Figure~\ref{fig:impurity} shows some possible defects;
the corresponding perturbations are
\begin{eqnarray}
\delta H^{(1)} &\sim& \vec{S}(x_0) \cdot \vec{S}(x_0+1)
\sim B^{(1)} (x_0) ~, \\
\delta H^{(2)} &\sim& \vec{S}(x_0 - 1) \cdot \vec{S}(x_0 + 1)
\sim B^{(2)} (x_0 - 1) ~, \\
\delta H^{(3)} &\sim&
\vec{S}(x_0) \cdot \left[ \vec{S}(x_0 - 1) + \vec{S}(x_0 + 1) \right] \\
&\sim& B^{(1)}(x_0 - 1) + B^{(1)}(x_0) ~, \\
\delta H^{(4)} &\sim&
\vec{S}(x_0) \cdot \left[ \vec{S}(x_0 - 2) + \vec{S}(x_0 + 2) \right] \\
&\sim& B^{(2)}(x_0 - 2) + B^{(2)}(x_0) ~.
\end{eqnarray}
Here $B^{(1)}$ and $B^{(2)}$ are given by Eq.~(\ref{Bn}).
We can characterize the defects by symmetry.  In the 1D chain picture,
$\delta H^{(1)}$ represents defects symmetric under inversion in a
bond center, while $\delta H^{(2,3,4)}$ are defects symmetric
under inversion in a site.  One can readily check that
$\delta H^{(2,3,4)}$ give equivalent expressions up to constant
factors and, importantly, contain all $Q$ modes in general.
We see that although the defects can be characterized as two
distinct symmetry types $\delta H^{(1)}$ and $\delta H^{(2)}$,
the perturbations to the Hamiltonian have the same dynamical field
content and differ only by constant phases.
This is unlike the Bethe phase of the 1D Heisenberg chain where a
bond-symmetric perturbation contains a relevant contribution from
a $Q=\pi$ bond operator while a site-symmetric perturbation does not.
\cite{Eggert92}

The scaling dimensions of the different contributions are
\begin{eqnarray}
\Delta[B_{2k_{Fa}}] &=& \frac{1}{2} + \frac{g}{4} ~,\\
\Delta[B_{4k_{F1}}] &=& g ~,\\
\Delta[B_{\pi/2}] &=& \frac{1}{2} + \frac{1}{4g} ~.
\end{eqnarray}
In the Spin Bose-Metal phase we have $g \leq 1$, so the $2k_{Fa}$ and
$4k_{F1}$ terms are always relevant 0+1D perturbations, while the
$\pi/2$ term is relevant if $g > 1/2$.
The relevant perturbations grow and one scenario is that they
eventually pin the fields at the origin.
Physically this leads to breaking the chain into two decoupled
semi-infinite systems, which we can then study separately.
The pinning conditions on the fields at the defect can be guessed by
considering the most relevant perturbation and minimizing the
corresponding energy.
We expect the $B_{2k_{Fa}}$ and $B_{4k_{F1}}$ terms to be the dominant,
which would
\begin{equation}
{\rm Pin}~\theta_{1\sigma}(x_0),~\theta_{2\sigma}(x_0),~\theta_{\rho-}(x_0) ~.
\label{pin}
\end{equation}
This is the case that we focus on.  In Appendix~\ref{app:pinII}
we will consider pinning conditions preferred by the $B_{\pi/2}$ term,
which may be of interest in the borderline case $g=1$.

A comment is in order.  On physical grounds, the symmetry of the
defect perturbation is important.
For the case with no site inversion symmetry like the impurity 1) in
Fig.~\ref{fig:impurity}, we can envision a possible outcome of the
RG growth of the perturbation by considering a situation where the
defect bond is strong.  The two spins will form a singlet, and if we
integrate it out, we get two semi-infinite chains weakly coupled to
each other, which under further RG will eventually flow to decoupled
semi-infinite systems with pinned values of the fields at the boundary.
We can envision more general situations where an even number of
spins will form a strongly coupled cluster with a singlet ground state,
and upon integrating this out we again have two weakly coupled
semi-infinite systems.
Below, we will consider a fixed point of a semi-infinite system and
give physical calculations of the bond textures and the oscillating
susceptibility near the boundary (impurity).
Turning to the case with impurities with site inversion symmetry
like 2), 3), 4) in Fig.~\ref{fig:impurity}, such reasoning
would give us a half-integer spin (formed by some effective strongly
coupled cluster with an odd number of sites) weakly coupled to two
semi-infinite systems.  This would need to be analyzed further,
which we briefly discuss in Sec.~\ref{subsec:other}.

\subsection{Physical calculations of oscillating susceptibility and
bond textures in the fixed-point theory of semi-infinite chains}
\label{subsec:physical}

From now on, we set the location of the defect to be the origin.
We work with a semi-infinite system with specified boundary conditions
at the origin and calculate the bond energy texture
\begin{eqnarray}
\la B(x) \ra = \sum_{a=1,2} \la B_{2k_{Fa}}(x) \ra
+ \la B_{4k_{F1}}(x) \ra + \la B_{\frac{\pi}{2}}(x) \ra ~.
\label{bondtexture}
\end{eqnarray}
We also calculate the local spin susceptibility, which can be measured in
Knight shift experiments.  We will see that there are contributions that
oscillate as a function of distance from the boundary:
$\chi(x) = \chi^{uni}(x) + \chi^{osc}(x)$; in fact, $\chi^{osc}(x)$
dominates over $\chi^{uni}(x)$ and can produce strong inhomogeneous
broadening of the NMR lineshapes.
The local spin susceptibility $\chi(i)$ at a lattice site $i$ measured
in a small uniform magnetic field $h$ is
\begin{eqnarray}
\chi(i) \equiv \frac{\partial \la S_i^z \ra}{\partial h} \Big|_{h=0}
= \beta \la S_i^z S_{\rm tot}^z \ra ~,
\end{eqnarray}
where $S_{\rm tot}^z \equiv \sum_j S_j^z$ is the total spin and
$\beta$ is the inverse temperature.
Rewriting the spin operators in terms of bosonic fields introduced above,
\begin{eqnarray}
&& \chi^{osc}(x) = \beta
\left\la S^z_{osc}(x) \int_0^\infty dy S^z_{uni}(y) \right\ra ~,
\end{eqnarray}
where $S^z_{osc} = \sum_Q e^{i Q x} S^{z}_Q$ and we are interested in
$Q = 2k_{F1}$, $2k_{F2}$, $\pi/2$, and $\pi$; while
$S^z_{uni}(y) = \sum_{a=1,2} \frac{\partial_y \theta_{a\sigma}(y)}{\sqrt{2}\pi}$.
Hence we define
\begin{eqnarray}
\chi^{osc}_Q \equiv \beta
\la e^{i Q x} S^z_Q(x) \int_0^\infty dy S^z_{uni}(y) + {\rm c.c.} \ra ~.
\end{eqnarray}

We consider the pinning Eq.~(\ref{pin}) driven by the relevant local
terms $B_{2k_{Fa}}, B_{4k_{F1}}$; in order to minimize these energies,
the natural pinning values of $\theta_{1\sigma}(0)$ and
$\theta_{2\sigma}(0)$ are
\begin{eqnarray}
\cos[\sqrt{2} \theta_{a\sigma}(0)] = \pm 1
\Rightarrow \sqrt{2} \theta_{a\sigma}(0) = {\rm integer} \times \pi ~.
\end{eqnarray}
The pinning value of the field $\theta_{\rho-}$ depends on the details
such as the amplitudes and phases $\gamma$ in
Eqs.~(\ref{bondenergy1})-(\ref{bondenergy3}).
As discussed in Appendix~\ref{app:formulas}, the pinning of a $\theta$
at the origin implies stronger fluctuation of the dual field $\varphi$
and consequently $\la e^{i \varphi(x)} \ra = 0$.

\underline{Bond energy texture} is given by Eq.~(\ref{bondtexture}).
The $\la B_{\pi/2}(x) \ra$ term vanishes and the other contributions
can be easily derived by applying the formulas in
Appendix~\ref{app:formulas}:
\begin{eqnarray}
\label{B_2kF}
\la B_{2k_{Fa}}(x) \ra &\simeq&
\frac{ A_{2k_{Fa}} \cos(2k_{Fa} x + \delta_{2k_{Fa}}) }
{ \left[\frac{v_a \beta}{\pi} \sinh(\frac{2\pi x}{v_a \beta})
     \right]^{\frac{1}{2}}
  \left[\frac{v \beta}{\pi} \sinh(\frac{2\pi x}{v \beta})
     \right]^{\frac{g}{4}} } ~,\\
\label{B_4kF}
\la B_{4k_{F1}}(x) \ra &\simeq&
\frac{ A_{4k_{F1}} \cos(4k_{F1} x + \delta_{4k_{F1}}) }
     { \left[\frac{v \beta}{\pi} \sinh(\frac{2\pi x}{v \beta}) \right]^g } ~,
\end{eqnarray}
where $a=1,2$; $A_Q$ are some amplitudes; and $\delta_Q$ are phases that
depend on the pinned $\theta$ values at the origin and are ultimately
determined by the details of the defect.
At low temperature $T \to 0$, we have the following behavior as a
function of the distance $x$ from the open boundary (defect):
\begin{eqnarray}
\la B_{2k_{Fa}}(x) \ra &\sim&
\frac{\cos(2k_{Fa} x + \delta_{2k_{Fa}})}{x^{\frac{1}{2} + \frac{g}{4}}}~,\\
\la B_{4k_{F1}}(x) \ra &\sim&
\frac{\cos(4k_{F1} x + \delta_{4k_{F1}})}{x^g} ~.
\end{eqnarray}
Thus, at low temperature the bond energy texture around the impurity
reveals the correlations present in the system, and the physics can be
viewed as a ``nucleation'' of the dominant ``bond orders'' near the
defect.  If we can tune the Luttinger parameter $g$, we see that
there are two regimes: for $2/3 < g < 1$ the $2k_{Fa}$ terms dominate,
while for $g < 2/3$ the $4k_{F1}$ dominates.

Turning to the \underline{oscillating susceptibility}, the
$\chi^{osc}_{\pi/2}$ term vanishes and only the $\chi^{osc}_{2k_{Fa}}$
and $\chi^{osc}_\pi$ contribute to the final result.
Applying the formulas from Appendix~\ref{app:formulas} gives
\begin{eqnarray}
\label{chi_2kF}
\chi^{osc}_{2k_{Fa}} &\simeq&
\frac{ C_{2k_{Fa}} \cdot x \cdot \cos(2k_{Fa} x + \delta_{2k_{Fa}}^\prime) }
     { \left[ \frac{v_a \beta}{\pi} \sinh\left( \frac{2\pi x}{v_a \beta}\right)
         \right]^{\frac{1}{2}}
       \left[ \frac{v \beta}{\pi} \sinh\left( \frac{2\pi x}{v \beta} \right)
         \right]^{\frac{g}{4}} } ~,\\
\label{chi_pi}
\chi^{osc}_\pi &\simeq&
\frac{ C_\pi \cdot x \cdot (-1)^x }
{ \left[ \frac{v_1 \beta}{\pi} \sinh\left(\frac{2\pi x}{v_1 \beta}\right)
    \right]^{\frac{1}{2}}
  \left[ \frac{v_2 \beta}{\pi} \sinh\left(\frac{2\pi x}{v_2 \beta}\right)
    \right]^{\frac{1}{2}} } ~,
\end{eqnarray}
where $a=1,2$; $C_Q$ are some constant amplitudes; and $\delta_Q^\prime$
some phases absorbing all pinned field values and eventually determined
by the details of the defect.
At low temperatures $T \to 0$, the oscillating susceptibilities at
$2k_{Fa}$ and $\pi$ become
\begin{eqnarray}
\chi^{osc}_{2k_{Fa}} (x) &\sim& x^{\frac{1}{2} - \frac{g}{4}}
\cos{(2k_{Fa} x + \delta_{2k_{Fa}}^\prime)} ~, \\
\chi^{osc}_\pi (x) &\sim& x^0 (-1)^x ~.
\end{eqnarray}
The envelope function in the first line satisfies
$x^{\frac{1}{2} - \frac{g}{4}} \geq x^{\frac{1}{4}}$,
which comes from the condition $g < 1$.
Therefore, at low temperatures the oscillating susceptibility at
$2k_{Fa}$ actually {\it increases} with the distance from the open end.
On the other hand, the oscillating susceptibility at $\pi$ reaches a
constant amplitude.

To conclude the discussion of the semi-infinite system with the
boundary conditions Eq.~(\ref{pin}), we note that this fixed point
is stable [e.g., the scaling dimension of $B_{\pi/2}(0)$ becomes
$1/2 + 1/(2g) > 1$, so it is irrelevant].
The boundary spin operator has scaling dimension 1: e.g.,
$S^z_{\rm bound.} \sim \partial_x (\theta_{1\sigma} + \theta_{2\sigma})$
at the boundary.  Knowing the fixed-point theory of the semi-infinite
chain, we can briefly discuss other situations with impurities.
\cite{Eggert92, Rommer98, Rommer00, Eggert01}
(For a recent review of impurity problems, see
Ref.~\onlinecite{Affleck08}.)

\subsection{Other Situations with Impurities}
\label{subsec:other}

\subsubsection{Weakly coupled semi-infinite systems}
In this case, we imagine two semi-infinite chains coupled to each other
at the origin.  Since in each semi-infinite system the scaling dimension
of the boundary spin operator is $1$, the spin-spin coupling between the
two systems is irrelevant and they will decouple at low energies.
This is the reason why a non-magnetic impurity like 1) in
Fig.~\ref{fig:impurity} breaks the system into two halves at low energies
and the physical calculations in Sec.~\ref{subsec:physical} apply
generically.

\subsubsection{Spin-$\frac{1}{2}$ impurity coupled to a semi-infinite
system}
In this case, the spin-1/2 impurity is coupled to the boundary
spin operator which contains contributions from both ``$1\sigma$'' and
``$2\sigma$'' channels,
$\delta H = \lambda \vec{S}_{\rm imp} \cdot
(\vec{S}_{\rm bound., 1} + \vec{S}_{\rm bound., 2}) \to
\lambda_1 \vec{S}_{\rm imp} \cdot \vec{S}_{\rm bound., 1} +
\lambda_2 \vec{S}_{\rm imp} \cdot \vec{S}_{\rm bound., 2}$.
(The ``$\rho-$'' sector does not enter in the important terms.)
The couplings $\lambda_1$ and $\lambda_2$ are both marginal.
If they are marginally irrelevant, the impurity spin will decouple.
If one of the couplings is marginally relevant while the other is
marginally irrelevant, the relevant coupling will grow and the
impurity spin will be absorbed into the corresponding channel.
Finally, if both of the couplings are relevant, since the two channels
are not equivalent, one coupling will grow faster; a likely scenario is
that the impurity spin will be absorbed into the dominating channel and
eventually the two channels will decouple.

\subsubsection{Two semi-infinite systems coupled symmetrically to a
spin-$\frac{1}{2}$ impurity}
Now let us take two semi-infinite chains and couple them together
through a spin-1/2 impurity symmetrically.  This case is also relevant
for the site-symmetric non-magnetic impurities like impurity 2), 3),
and 4) in Fig.~\ref{fig:impurity}:  The reason is that because of the
site inversion symmetry, the non-magnetic impurity affects an even number
of bonds which couple an odd number of spins; then we can imagine a
strongly-coupled cluster with the odd number of spins, which will
effectively behave as a half-integer spin weakly coupled to the
left and right semi-infinite systems.

The situation is more complex than in the previous subsection because
we now have symmetry between the two semi-infinite systems, reminiscent
of the 2-channel Kondo problem.  We can imagine the following
possibilities.  When all couplings are marginally irrelevant,
the impurity spin and the two semi-infinite systems will decouple at
low energies (and the physical calculations of textures in
Sec.~\ref{subsec:physical} are valid in this case).
Suppose now we have marginally relevant couplings and the dominant
growth is for the channels $1\sigma$ in the two semi-infinite systems.
One is tempted to speculate about the possibility of ``healing'' the channels
$1\sigma$ across the impurity, while the channels $2\sigma$ remain open.
However, it is likely that this is not a stable fixed point in the
presence of the allowed terms in the Hamiltonian coming from the
microscopic ladder system.  While the eventual outcome is not clear
and depends on details, on physical grounds we again expect arriving at
some stage at a fixed point with some odd number of spins forming a
half-integer spin that is decoupled from two semi-infinite systems.

\section{Conclusions}
To summarize, following the theoretical description\cite{Sheng09} of the
Spin Bose-Metal phase in the triangular strip spin-1/2 model with ring
exchanges, we discussed the effects due to different types of impurities.
The defects can have additional bond or site symmetry in the 1D zigzag
chain language.  We first treated the defects as local perturbations in
the Hamiltonian and saw that all types produce relevant perturbations,
eventually breaking the system into two halves and a separate decoupled
cluster of spins.  In the bond-symmetric case (or more general cases
with no symmetries) the decoupled cluster is likely to be non-magnetic,
while in the site-symmetric case it has half-integer spin,
and the details of such fixed points depend on the microscopic
details.\cite{Eggert92, Rommer00, Eggert01}
This analysis also motivated appropriate boundary conditions for
pinning the fields in the fixed point theory for the semi-infinite
systems.

For such a semi-infinite chain, we calculated the bond energy texture
near the boundary and found power law decays
Eqs.~(\ref{B_2kF}, \ref{B_4kF}) of the oscillating components at
wavevectors $2k_{Fa}$, $4k_{F1}$.
The dominant power law switches from the $2k_{Fa}$ to the $4k_{F1}$
when the Luttinger parameter $g$ drops below $2/3$.
We suggest that characterizing such bond textures in numerical
studies, e.g. DMRG,\cite{Sheng09} could be useful for determining the Luttinger
parameter $g$ of the SBM theory.

We also calculated the oscillating susceptibilities at $2k_{Fa}$ and
$\pi$, Eqs.~(\ref{chi_2kF}, \ref{chi_pi}), which behave differently at
low temperatures.
The susceptibilities at $2k_{Fa}$ actually {\it increase} with the
distance from the boundary in the limit of zero temperature
(and zero field), while the susceptibility at $\pi$ becomes
distance-independent. Transfer-Matrix density-matrix Renormalization Group (TMRG)
\cite{Rommer98, Rommer00, TMRG96} technique can measure local susceptibility at finite 
temperature and can be useful for exploring the susceptibility near defects in numerical studies.
The rate of increase at $2k_{Fa}$ is slower than in the 1D chain,\cite{Eggert95}
but would still produce strong NMR line broadening at low temperatures. Of course,
this is the result for the long-distance behavior along the 1D direction.
If we are thinking about the 2D spin liquid, we would likely expect a power-law
decay away from an impurity.\cite{Kim03, Kolezhuk06, Gregor08} Nevertheless,
the persistence of the oscillating susceptibilities on the quas-1D ladders suggests
that in the 2D case the decay may be slow and also produce significant inhomogeneous
line broadening. Finally, in this paper, we focused on non-magnetic impurities and 
the simplest ``fixed-point'' model with open boundary. We have not touched interesting 
and experimentally relevant crossovers present for a magnetic impurity weakly coupled 
to the system.\cite{Rommer98, Rommer00}  Here again theoretical and numerical studies 
similar to Ref.~\onlinecite{Rommer98, Rommer00} could be very helpful, for example, 
in estimating the size of the Kondo screening cloud, which is an additional and potentially 
large effect near the magnetic impurity.

\acknowledgments

We would like to thank D.~N.~Sheng and M.~P.~A.~Fisher for collaborations
and discussions.
This research is supported by the A.~P.~Sloan Foundation.

\appendix

\section{One mode theory on a semi-infinite chain}
\label{app:formulas}

For the simplest case,\cite{Eggert92, Eggert95, Rommer00, Bortz05}
consider a one mode theory on a semi-infinite chain with pinned value
at the origin, $\theta(0, \tau) = pinned = \theta_0$.
The action is
\begin{eqnarray}
S = \int_0^\infty dx \int_0^\beta d\tau \frac{1}{2\pi g}
\left[ v (\partial_x \theta)^2 + \frac{1}{v} (\partial_\tau \theta)^2 \right] ~.
\end{eqnarray}
The correlation functions needed in this paper are
\begin{eqnarray}
&& \hspace{-8mm}
\left\langle e^{i u \theta(x,\tau)} \right\rangle
\simeq  \frac{A e^{i u \theta_0}}{[\frac{v \beta}{\pi} \sinh{(\frac{2\pi x}{v \beta})}]^{\frac{u^2 g}{4}}} ~,\\
&& \hspace{-8mm}
\left\langle e^{i u \theta(x,\tau)}
             \int_0^\infty dy \frac{\partial_y \theta(y,\tau')}{\pi}
\right\rangle
= \frac{i u g x}{\beta v} \times
\left\langle e^{i u \theta(x,\tau)} \right\rangle \\
&&\hspace{25mm} \simeq \frac{i u g x}{\beta v} \frac{A e^{i u \theta_0}}{[\frac{v \beta}{\pi} \sinh{(\frac{2\pi x}{v \beta})}]^{\frac{u^2 g}{4}}} ~, \\
&& \hspace{-8mm}
\left\langle e^{i u \varphi(x,\tau)} \right\rangle = 0 ~.
\end{eqnarray}
Here $u$ is a parameter depending on which quantity is being measured and
$A$ is some real constant.  The $\la e^{i u \theta} \ra$ is non-zero
because of the pinning at $x=0$ and decays as a power-law away from the
origin at $T=0$.  On the other hand, the conjugate field $\varphi$
fluctuates more strongly than in the bulk and $\la e^{i u \varphi} \ra = 0$
everywhere.

We can similarly consider a one mode theory with the dual field
pinned at the origin, $\varphi(0,\tau) = pinned = \varphi_0$.
It is convenient to work with the action
\begin{eqnarray}
S = \int_0^\infty dx \int_0^\beta d\tau \frac{g}{2\pi}
\left[ v (\partial_x \varphi)^2 + \frac{1}{v} (\partial_\tau \varphi)^2 \right] ~.
\end{eqnarray}
The correlation function needed is
\begin{eqnarray}
&& \left\langle e^{i u \varphi(x,\tau)} \right\rangle
\simeq  \frac{\tilde{A} e^{i u \varphi_0}}{[\frac{v \beta}{\pi} \sinh{(\frac{2\pi x}{v \beta})}]^{\frac{u^2}{4g}}} ~,
\end{eqnarray}
where $\tilde{A}$ is some constant.

\section{Calculations in a fixed point theory of a semi-infinite
system with pinned $\theta_{\sigma+}(0)$, $\varphi_{\sigma-}(0)$, and
$\varphi_{\rho-}(0)$}
\label{app:pinII}

Here we consider the theory Eq.~(\ref{LSBM}) on a semi-infinite chain
with boundary conditions
\begin{equation}
{\rm Pin}~\theta_{\sigma+}(0),~\varphi_{\sigma-}(0),~\varphi_{\rho-}(0) ~.
\label{pinII}
\end{equation}
This can arise if we minimize the $B_{\pi/2}(0)$ perturbation
instead of the $B_{2k_{Fa}}(0)$ and $B_{4k_{F1}}(0)$, see
Eqs.~(\ref{bondenergy1}-\ref{bondenergy3}).
Coming from the microscopic ladder spin system in the SBM phase
with $g<1$, this fixed point is unstable to the allowed $B_{2k_{Fa}}(0)$
terms.  Nevertheless, it can be of interest in the
special case with $g=1$, which is realized, e.g., by the Gutzwiller
wavefunctions or at phase transitions out of the SBM.\cite{Sheng09}
The calculations of the physical textures are simple and we summarize
these below.

Because of the pinning Eq.~(\ref{pinII}), the dual fields
$\varphi_{\sigma+}$, $\theta_{\sigma-}$, and $\theta_{\rho-}$ fluctuate
more strongly.  The only non-vanishing term in the
\underline{bond energy texture} is
\begin{eqnarray}
\la B(x) \ra = \la B_{\pi/2}(x) \ra ~.
\end{eqnarray}
If $v_1 = v_2$, then ``$\sigma+$'' and ``$\sigma-$'' variables decouple
and we can apply the formulas in Appendix~\ref{app:formulas}.
In the general case $v_1 \neq v_2$, the calculations are more demanding
but the result is simple:
\begin{eqnarray}
\langle B_{\pi/2}(x) \rangle \simeq
\frac{ A_{\pi/2} \cos{(\frac{\pi}{2}x + \delta_{\pi/2})} }
     { \left[ \frac{v_r \beta}{\pi} \sinh{(\frac{2\pi x}{v_r \beta})}
         \right]^{\frac{1}{2}}
       \left[ \frac{v \beta}{\pi} \sinh{(\frac{2\pi x}{v \beta})}
         \right]^{\frac{1}{4g}} } ~,
\end{eqnarray}
where
$\frac{1}{v_r} = \frac{1}{2} \left(\frac{1}{v_1} + \frac{1}{v_2}\right)$,
$A_{\pi/2}$ is some amplitude and $\delta_{\pi/2}$ is a constant phase.
The pinned values at the origin as well as the Klein numbers enter
in the same way as they enter the assumed minimization of $B_{\pi/2}(0)$,
Eq.~(\ref{bondenergy3}), so the final result depends only on the
physical details of this term at the origin.
In the limit $T \to 0$,
\begin{eqnarray}
\langle B_{\pi/2}(x) \rangle \sim
\frac{\cos{(\frac{\pi}{2}x + \delta_{\pi/2})}}{x^{\frac{1}{2}+\frac{1}{4g}}} ~.
\end{eqnarray}

As for the \underline{oscillating susceptibility}, similarly to the
bond energy texture, only the $Q=\pi/2$ term contributes to the
final result.  Again, for $v_1 = v_2$ we can apply the formulas
in Appendix~\ref{app:formulas}, while in the general case we get
\begin{eqnarray}
\chi^{osc}_{\pi/2} \simeq
\frac{ C \cdot x \cdot \cos(\frac{\pi}{2}x + \delta_{\pi/2}^\prime) }
     { \left[ \frac{v_r \beta}{\pi} \sinh(\frac{2\pi x}{v_r \beta})
         \right]^{\frac{1}{2}}
       \left[ \frac{v \beta}{\pi} \sinh(\frac{2\pi x}{v \beta})
         \right]^{\frac{1}{4g}} } ~,
\end{eqnarray}
where $C_{\pi/2}$ is some amplitude and $\delta_{\pi/2}^\prime$ is a
constant phase absorbing all pinned values and eventually determined by
the details of the defining energy $B_{\pi/2}(0)$.
In the low temperature limit $T \to 0$,
\begin{eqnarray}
\chi^{osc}_{\pi/2} \sim x^{\frac{1}{2} - \frac{1}{4g}}
\cos(\frac{\pi}{2}x + \delta_{\pi/2}^\prime) ~.
\end{eqnarray}


\end{document}